\documentclass[12pt]{article}
\usepackage{epsf}
\title{ {\bf
$b\rightarrow s \gamma$ decay in the two Higgs doublet model with flavor 
changing neutral currents}}
\author{\vspace{1cm}\\
         {\bf T. M. Aliev} , \\
        Physics Department, Middle East Technical University \\
        Ankara, Turkey\\
        \vspace{5mm}\\
        {\bf E. O. Iltan}
        \thanks{E-mail address:
        eiltan@heraklit.physics.metu.edu.tr}
 \\
        Physics Department, Middle East Technical University \\
        Ankara, Turkey\\}

\date{}

\begin{document}
\setlength{\baselineskip}{24pt}
\maketitle
\setlength{\baselineskip}{7mm}
\begin{abstract}
We study the $b\rightarrow s\gamma$ decay including the next to leading 
QCD corrections in the two Higgs doublet model with flavor changing neutral 
currents at the tree level.
We find the constraints to the flavor
changing parameters of the model, using the experimental results
on the branching ratio of $B \rightarrow X_s \gamma$ decay, provided by 
the CLEO Collaboration, the restrictions coming from the 
$\Delta F=2$ ($F=K,D,B$) mixing and the $\rho$ parameter.
\end{abstract} 
\thispagestyle{empty}
\newpage
\setcounter{page}{1}

\section{Introduction}
Rare B meson decays are one of the most promising research area in particle
physics and lie on the focus of theoretical and experimental physicists.
In the Standard Model (SM), they are induced by flavor changing neutral 
currents (FCNC) at loop level and therefore sensitive to the gauge structure 
of the theory. From the experimental point of view, they play an outstanding
role in the precise determination of the fundamental parameters of the SM, 
such as Cabbibo-Kobayashi-Maskawa (CKM) matrix elements, leptonic decay
constants, etc. Furthermore, these decays provide a sensitive 
test to the new physics beyond the SM, such as
two Higgs Doublet model (2HDM), Minimal Supersymmetric extension of the SM
(MSSM) \cite{hewett}, etc. 
Among the rare B decays, $b\rightarrow s\gamma$ has received considerable
interest since the branching ratios (Br) of the inclusive $B\rightarrow X_s
\gamma$ \cite{cleo} and exclusive $B\rightarrow K^* \gamma $ \cite{rammar} 
have been already measured experimentally. Recently,
the new experimental results for the inclusive $b\rightarrow s\gamma$ decay 
are announced by CLEO and ALEPH Collaborations \cite{cleo2}. 
Therefore, the $b\rightarrow s\gamma$ decay is under an extensive investigation
in the framework of various extensions of the SM, in order to get
information about the model parameters or improve the existing restrictions.

It is well known that the FCNC are forbidden at the tree level in the SM. This 
restriction is achieved in the extended model with the additional conditions.
2HDM is one of the simplest extensions of the SM, obtained by the 
addition of a new scalar $SU(2)$ doublet. The Yukawa lagrangian 
causes that the model possesses phenemologicaly dangerous FCNC's at 
the tree level. To protect the model from such terms, 
the ad hoc discrete symmetry \cite{weinberg} on the 2HDM scalar potential 
and the Yukawa interaction is proposed and 
there appear two different versions of the 2HDM depending on whether up and 
down quarks couple to the same or different scalar doublets. 
In model I, the up and down quarks get mass via vacuum
expectation value (v.e.v.) of only one Higgs field. In model II, which 
coincides with the MSSM in the 
Higgs sector, the up and down quarks get mass via v.e.v. of the Higgs fields 
$H_{1}$ and $H_{2}$ respectively where $H_{1} (H_{2})$ corresponds to first 
(second) Higgs doublet of 2HDM \cite{kane}. In the absence of the mentioned 
discrete symmetry, FCNC appears at the tree level and this model is called
as model III in current literature \cite{atwood1,atwood2,soniref}.
A comprehensive phenemological analysis of the model III was done in series of
works \cite{atwood1,atwood2,soni2}. 
In particular, from a purely phenomenological point of view, low energy
experiments involving $K^0 - \bar{K}^0$, $B^0 - \bar{B}^0$,
$K_{L}\rightarrow \mu\bar{\mu}$, etc, place strong constraints on the existence of
tree level flavor changing (FC) transitions, existing in the model III.
  
In the present work, we examine the $b\rightarrow s\gamma$ decay in the
model III, taking the next to leading (NLO) QCD corrections into 
account, in a more detailed analysis compared to one existing in literature 
(see \cite{atwood1,soni2}). 
Further, we obtain the constraints for the neutral couplings 
$\xi_{N tt}^{U}$, $\xi_{N bb}^{D}$ and $\xi_{N tc}^{U}$ with the assumption 
that $\xi_{N cc}^{U}$, $\xi_{N sb}^{D}$, $\xi_{N ss}^{D}$ and 
the other couplings which include the first generation indices 
are negligible compared to former ones (for the definition of $\xi_{N,ij}$
see section 2). Our predictions are based on the CLEO measurement 
$B\rightarrow X_s \gamma$ and the restrictions coming from the 
$\Delta F=2$ ($F=K,D,B$) mixing and the $\rho$ parameter \cite{soni2}.
Note that NLO QCD corrections to the $b\rightarrow s\gamma$ decay in  
2HDM (for model I and II) were calculated in \cite{greub2,ciuchini2}.

The paper is organized as follows:
In Section 2, we present the NLO QCD corrected Hamiltonian responsible for
the $b\rightarrow s\gamma$ decay in the model III and discuss the effects
of the additional $Left-Right$ flipped operators to the decay rate. 
Section 3 is devoted to the constraint analysis, more precisely 
to the the ratios $\frac{\xi_{N tt}^{U}}{\xi_{N bb}^{D}}$, 
$\frac{\xi_{N tc}^{U}}{\xi_{N tt}^{U}}$ and our conclusions.

\section{\bf Next to leading improved short-distance contributions in 
the model III for the decay $b\rightarrow s \gamma $ }
Before presenting the NLO  QCD corrections to the 
$b\rightarrow s \gamma $ decay amplitude in the 2HDM (model III), 
we would like to remind briefly the main features of the 2HDM.
The Yukawa interaction for the general case is
\begin{eqnarray}
{\cal{L}}_{Y}=\eta^{U}_{ij} \bar{Q}_{i L} \tilde{\phi_{1}} U_{j R}+
\eta^{D}_{ij} \bar{Q}_{i L} \phi_{1} D_{j R}+
\xi^{U}_{ij} \bar{Q}_{i L} \tilde{\phi_{2}} U_{j R}+
\xi^{D}_{ij} \bar{Q}_{i L} \phi_{2} D_{j R} + h.c. \,\,\,
\label{lagrangian}
\end{eqnarray}
where $L$ and $R$ denote chiral projections $L(R)=1/2(1\mp \gamma_5)$,
$\phi_{i}$ for $i=1,2$, are the two scalar doublets, $\eta^{U,D}_{ij}$
and $\xi^{U,D}_{ij}$ are the matrices of the Yukawa couplings.
For convenience we choose $\phi_1$ and $\phi_2$ in the following basis:
\begin{eqnarray}
\phi_{1}=\frac{1}{\sqrt{2}}\left[\left(\begin{array}{c c} 
0\\v+H^{0}\end{array}\right)\; + \left(\begin{array}{c c} 
\sqrt{2} \chi^{+}\\ i \chi^{0}\end{array}\right) \right]\, ; 
\phi_{2}=\frac{1}{\sqrt{2}}\left(\begin{array}{c c} 
\sqrt{2} H^{+}\\ H_1+i H_2 \end{array}\right) \,\, ,
\label{choice}
\end{eqnarray}
where the vacuum expectation values are,  
\begin{eqnarray}
<\phi_{1}>=\frac{1}{\sqrt{2}}\left(\begin{array}{c c} 
0\\v\end{array}\right) \,  \, ; 
<\phi_{2}>=0 \,\, .
\label{choice2}
\end{eqnarray}
This choice permits us to write the FC part of the 
interaction as 
\begin{eqnarray}
{\cal{L}}_{Y,FC}=
\xi^{U}_{ij} \bar{Q}_{i L} \tilde{\phi_{2}} U_{j R}+
\xi^{D}_{ij} \bar{Q}_{i L} \phi_{2} D_{j R} + h.c. \,\, ,
\label{lagrangianFC}
\end{eqnarray}
with the following advantages: 
\begin{itemize}
\item doublet $\phi_{1}$ corresponds to the scalar doublet of the SM and 
$H_{0}$ to the SM Higgs field. This part of the Yukawa Lagrangian is
responsible for the generation of the fermion masses with the couplings 
$\eta^{U,D}$.

\item all new scalar fields belong to the $\phi_{2}$ scalar doublet.
\end{itemize}
The couplings  $\xi^{U,D}$ are the open window for the tree level FCNC's
and can be expressed for the FC charged interactions as
\begin{eqnarray}
\xi^{U}_{ch}&=& \xi_{neutral} \,\, V_{CKM} \nonumber \,\, ,\\
\xi^{D}_{ch}&=& V_{CKM} \,\, \xi_{neutral} \,\, ,
\label{ksi1} 
\end{eqnarray}
where $\xi^{U,D}_{neutral}$ 
\footnote{In all next discussion we denote $\xi^{U,D}_{neutral}$ 
as $\xi^{U,D}_{N}$.} 
is defined by the expression
\begin{eqnarray}
\xi^{U,D}_{N}=(V_L^{U,D})^{-1} \xi^{U,D} V_R^{U,D}\,\, .
\label{ksineut}
\end{eqnarray}
Here the charged couplings appear as a linear combinations of neutral 
couplings multiplied by $V_{CKM}$ matrix elements. This gives an important
distinction between model III and model II (I).

After this preliminary remark, let us discuss the NLO QCD corrections to
the $b\rightarrow s\gamma$ decay in the 2HDM for the general case. The 
appropriate framework is that of an effective theory obtained by integrating 
out the heavy degrees of freedom, which are, in this context,
$t$ quark, $W^{\pm}, H^{\pm}, H_{1}$, and $H_{2}$ bosons, where $H^{\pm}$ 
denote charged ,$H_{1}$ and $H_{2}$ denote neutral Higgs bosons. 
The LLog QCD corrections are done through matching the full theory with the
effective low energy theory at the high scale $\mu=m_{W}$ and 
evaluating the Wilson coefficients from $m_{W}$ down to the lower scale $\mu\sim O(m_{b})$.
Note that we choose the higher scale as $\mu=m_{W}$ since the evaluation 
from the scale $\mu=m_{H^{\pm}}$ to $\mu=m_{W}$ gives negligible
contribution to the Wilson coefficients. Here we assume that the charged
Higgs boson is heavy due to theoretical analysis of the 
$b\rightarrow s\gamma$ decay (see \cite{greub2, ahi}).
      
The effective Hamiltonian relevant for $b\rightarrow s \gamma$ decay is
\begin{eqnarray}
{\cal{H}}_{eff}=-4 \frac{G_{F}}{\sqrt{2}} V_{tb} V^{*}_{ts} 
\sum_{i}C_{i}(\mu) O_{i}(\mu) \, \, ,
\label{hamilton}
\end{eqnarray}
where the $O_{i}$ are operators given in eq.~(\ref{o7}) 
and the $C_{i}$ are Wilson coefficients
renormalized at the scale $\mu$. The coefficients are calculated 
perturbatively and expressed as functions of the heavy particle masses
in the theory.

The operator basis depends on the model used and 
the conventional choice in the case of SM, 2HDM model II (I) and MSSM is
\begin{eqnarray}
 O_1 &=& (\bar{s}_{L \alpha} \gamma_\mu c_{L \beta})
               (\bar{c}_{L \beta} \gamma^\mu b_{L \alpha}), \nonumber   \\
 O_2 &=& (\bar{s}_{L \alpha} \gamma_\mu c_{L \alpha})
               (\bar{c}_{L \beta} \gamma^\mu b_{L \beta}),  \nonumber   \\
 O_3 &=& (\bar{s}_{L \alpha} \gamma_\mu b_{L \alpha})
               \sum_{q=u,d,s,c,b}
               (\bar{q}_{L \beta} \gamma^\mu q_{L \beta}),  \nonumber   \\
 O_4 &=& (\bar{s}_{L \alpha} \gamma_\mu b_{L \beta})
                \sum_{q=u,d,s,c,b}
               (\bar{q}_{L \beta} \gamma^\mu q_{L \alpha}),   \nonumber  \\
 O_5 &=& (\bar{s}_{L \alpha} \gamma_\mu b_{L \alpha})
               \sum_{q=u,d,s,c,b}
               (\bar{q}_{R \beta} \gamma^\mu q_{R \beta}),   \nonumber  \\
 O_6 &=& (\bar{s}_{L \alpha} \gamma_\mu b_{L \beta})
                \sum_{q=u,d,s,c,b}
               (\bar{q}_{R \beta} \gamma^\mu q_{R \alpha}),  \nonumber   \\  
 O_7 &=& \frac{e}{16 \pi^2}
          \bar{s}_{\alpha} \sigma_{\mu \nu} (m_b R + m_s L) b_{\alpha}
                {\cal{F}}^{\mu \nu},                             \nonumber       \\
 O_8 &=& \frac{g}{16 \pi^2}
    \bar{s}_{\alpha} T_{\alpha \beta}^a \sigma_{\mu \nu} (m_b R + m_s L)  
          b_{\beta} {\cal{G}}^{a \mu \nu},  
\label{o7}
\end{eqnarray}
where  
$\alpha$ and $\beta$ are $SU(3)$ colour indices and
${\cal{F}}^{\mu \nu}$ and ${\cal{G}}^{\mu \nu}$
are the field strength tensors of the electromagnetic and strong
interactions, respectively.

In our case, however, new operators with different chirality structures 
can arise since the general Yukawa lagrangian includes both $L$ and $R$ 
chiral interactions.
The conventional operator set is extended first adding two new
operators which are left-right analogues of $O_{1}$ and $O_{2}$, namely
\begin{eqnarray}
 O_9 &=& (\bar{s}_{L \alpha} \gamma_\mu c_{L \beta})
               (\bar{c}_{R \beta} \gamma^\mu b_{R \alpha}), \nonumber   \\
 O_{10} &=& (\bar{s}_{L \alpha} \gamma_\mu c_{L \alpha})
(\bar{c}_{R \beta} \gamma^\mu b_{R \beta}),
\label{newop1}
\end{eqnarray}
Further we need the second operator set $O'_{1} - O'_{10}$ which are 
flipped chirality partners of $O_{1} - O_{10}$:
\begin{eqnarray}
 O'_1 &=& (\bar{s}_{R \alpha} \gamma_\mu c_{R \beta})
               (\bar{c}_{R \beta} \gamma^\mu b_{R \alpha}), \nonumber   \\
 O'_2 &=& (\bar{s}_{R \alpha} \gamma_\mu c_{R \alpha})
               (\bar{c}_{R \beta} \gamma^\mu b_{R \beta}),  \nonumber   \\
 O'_3 &=& (\bar{s}_{R \alpha} \gamma_\mu b_{R \alpha})
               \sum_{q=u,d,s,c,b}
               (\bar{q}_{R \beta} \gamma^\mu q_{R \beta}),  \nonumber   \\
 O'_4 &=& (\bar{s}_{R \alpha} \gamma_\mu b_{R \beta})
                \sum_{q=u,d,s,c,b}
               (\bar{q}_{R \beta} \gamma^\mu q_{R \alpha}),   \nonumber  \\
 O'_5 &=& (\bar{s}_{R \alpha} \gamma_\mu b_{R \alpha})
               \sum_{q=u,d,s,c,b}
               (\bar{q}_{L \beta} \gamma^\mu q_{L \beta}),   \nonumber  \\
 O'_6 &=& (\bar{s}_{R \alpha} \gamma_\mu b_{R \beta})
                \sum_{q=u,d,s,c,b}
               (\bar{q}_{L \beta} \gamma^\mu q_{L \alpha}),  \nonumber   \\  
 O'_7 &=& \frac{e}{16 \pi^2}
          \bar{s}_{\alpha} \sigma_{\mu \nu} (m_b L + m_s R) b_{\alpha}
                {\cal{F}}^{\mu \nu},                             \nonumber       \\
 O'_8 &=& \frac{g}{16 \pi^2}
    \bar{s}_{\alpha} T_{\alpha \beta}^a \sigma_{\mu \nu} (m_b L + m_s R)  
          b_{\beta} {\cal{G}}^{a \mu \nu}, \nonumber \\ 
 O'_9 &=& (\bar{s}_{R \alpha} \gamma_\mu c_{R \beta})
               (\bar{c}_{L \beta} \gamma^\mu b_{L \alpha})\,\, , \nonumber   \\
 O'_{10} &=& (\bar{s}_{R \alpha} \gamma_\mu c_{R \alpha})
(\bar{c}_{L \beta} \gamma^\mu b_{L \beta})\,\,.
\label{newop2}
\end{eqnarray}
This extended basis is the same as the basis for 
$SU(2)_L\times SU(2)_R\times U(1)$ extensions of SM \cite{cho}.
Note that in the SM, model II (I) 2HDM and the MSSM,
the absence of $O_{7}'$ and $O_{8}'$ are a consequence of assumption 
$m_{s}/m_{b}\sim 0$.

In the calculations, we take only the charged Higgs contributions into 
account and neglect the effects of neutral Higgs bosons for the reasons 
given below:
The neutral bosons $H_0$, $H_1$ and $H_2$ are defined  in terms of the
mass eigenstates $\bar{H}_0$ ,$h_0$ and $A_0$ as 
\begin{eqnarray}
H_{0}&=&( \bar{H}_{0}cos \alpha - h_0 sin\alpha)+v \nonumber \, ,\\ 
H_{1}&=&( h_{0}cos \alpha + \bar{H}_0 sin\alpha) \nonumber \, ,\\ 
H_{2}&=&A_0 \,\,,
\label{neutrbos}
\end{eqnarray}
where $\alpha$ is the mixing angle and $v$ is proportional to the vacuum 
expectation value of the doublet $\phi_1$ (eq. (\ref{choice2})). Here we
assume that the massess of neutral Higgs bosons $h_0$ and $A_0$ are heavy
compared to the b-quark mass. The neutral Higgs scalar $h_0$ and pseduscalar 
$A_0$ give contribution only to $C_7$ for $b\rightarrow s\gamma$ decay. 
With the choice of $\alpha=0$, $C_7^{h_0}$ and $C_7^{A_0}$ can be expressed 
at $m_W$ level as
\begin{eqnarray}
C_7^{h_0}(m_W)&=& (V_{tb} V^{*}_{ts} )^{-1}\sum_{i=d,s,b} \bar{\xi}^{D}_{N,bi} 
\,\,\bar{\xi}^{D}_{N,is}\,  \frac{Q_i}{8\, m_i\, m_b}
\nonumber \,\,, \\ 
C_7^{A_0}(m_W)&=& (V_{tb} V^{*}_{ts} )^{-1}\sum_{i=d,s,b} \bar{\xi}^{D}_{N,bi} 
\,\, \bar{\xi}^{D}_{N,is}\, \frac{Q_i}{8\, m_i\, m_b}
\,\, ,
\label{c7A0h0}
\end{eqnarray}
where $m_i$ and $Q_i$ are the masses and charges of the down quarks 
($i=d,\,s,\,b$) respectively. Here we used the redefinition
\begin{eqnarray}
\xi^{U,D}=\sqrt{\frac{4 G_{F}}{\sqrt{2}}} \,\, \bar{\xi}^{U,D}\,\, .
\label{ksidefn}
\end{eqnarray} 
Eq. (\ref{c7A0h0}) shows that neutral Higgs bosons can give a large 
contribution to $C_7$, which does not respect the CLEO and ALEPH data 
\cite{cleo2}. 
At this stage we make an assumption that the couplings 
$\bar{\xi}^{D}_{N,is}$($i=d,s,b)$ and $\bar{\xi}^{D}_{N,db}$ are 
negligible to be able to reach the conditions 
$\bar{\xi}^{D}_{N,bb} \,\bar{\xi}^{D}_{N,is} <<1$ and 
$\bar{\xi}^{D}_{N,db} \,\bar{\xi}^{D}_{N,ds} <<1$.
These choices permit us to neglect the neutral Higgs effects.

Now, for the evaluation of Wilson coefficients, we need their initial values 
with standard matching computations. Denoting the Wilson coefficients for 
the additional charged Higgs contribution 
with $C_{i}^{H}(m_{W})$, we have the initial values of the Wilson
coefficients for the first set of operators (eqs.(\ref{o7}), (\ref{newop1}))  
\begin{eqnarray}
C^{H}_{1,\dots 6,9,10}(m_W)&=&0 \nonumber \, \, , \\
C_7^{H}(m_W)&=&\frac{1}{m_{t}^2} \,
(\bar{\xi}^{U}_{N,tt}+\bar{\xi}^{U}_{N,tc}
\frac{V_{cs}^{*}}{V_{ts}^{*}}) \, (\bar{\xi}^{U}_{N,tt}+\bar{\xi}^{U}_{N,tc}
\frac{V_{cb}}{V_{tb}}) F_{1}(y)\nonumber  \, \, , \\
&+&\frac{1}{m_t m_b} \, (\bar{\xi}^{U}_{N,tt}+\bar{\xi}^{U}_{N,tc}
\frac{V_{cs}^{*}}{V_{ts}^{*}}) \, (\bar{\xi}^{D}_{N,bb}+\bar{\xi}^{D}_{N,sb}
\frac{V_{ts}}{V_{tb}}) F_{2}(y)\nonumber  \, \, , \\
C_8^{H}(m_W)&=&\frac{1}{m_{t}^2} \,
(\bar{\xi}^{U}_{N,tt}+\bar{\xi}^{U}_{N,tc}
\frac{V_{cs}^{*}}{V_{ts}^{*}}) \, (\bar{\xi}^{U}_{N,tt}+\bar{\xi}^{U}_{N,tc}
\frac{V_{cb}}{V_{tb}})G_{1}(y)\nonumber  \, \, , \\
&+&\frac{1}{m_t m_b} \, (\bar{\xi}^{U}_{N,tt}+\bar{\xi}^{U}_{N,tc}
\frac{V_{cs}^{*}}{V_{ts}^{*}}) \, (\bar{\xi}^{D}_{N,bb}+\bar{\xi}^{U}_{N,sb}
\frac{V_{ts}}{V_{tb}}) G_{2}(y) \, \, .
\label{CoeffH}
\end{eqnarray}
The explicit forms of the Wilson coefficients in the SM ($C_{i}^{SM}(m_{W})$) 
is presented in the literature \cite{aburas}. 
For the primed Wison coefficients we get, 
\begin{eqnarray}
C^{\prime SM}_{1,\dots 6,9,10}(m_W)&=&0 \nonumber \, \, , \\
\label{Coeffsm2}
\end{eqnarray}
 
\begin{eqnarray}
C^{\prime H}_{1,\dots 6,9,10}(m_W)&=&0 \nonumber \, \, , \\
C^{\prime H}_7(m_W)&=&\frac{1}{m_t^2} \,
(\bar{\xi}^{D}_{N,bs}\frac{V_{tb}}{V_{ts}^{*}}+\bar{\xi}^{D}_{N,ss})
\, (\bar{\xi}^{D}_{N,bb}+\bar{\xi}^{D}_{N,sb}
\frac{V_{ts}}{V_{tb}}) F_{1}(y)\nonumber  \, \, , \\
&+& \frac{1}{m_t m_b}\, (\bar{\xi}^{D}_{N,bs}\frac{V_{tb}}{V_{ts}^{*}}
+\bar{\xi}^{D}_{N,ss}) \, (\bar{\xi}^{U}_{N,tt}+\bar{\xi}^{U}_{N,tc}
\frac{V_{cb}}{V_{tb}}) F_{2}(y)\nonumber  \, \, , \\
C^{\prime H}_8 (m_W)&=&\frac{1}{m_t^2} \,
(\bar{\xi}^{D}_{N,bs}\frac{V_{tb}}{V_{ts}^{*}}+\bar{\xi}^{D}_{N,ss})
\, (\bar{\xi}^{D}_{N,bb}+\bar{\xi}^{D}_{N,sb}
\frac{V_{ts}}{V_{tb}}) G_{1}(y)\nonumber  \, \, , \\
&+&\frac{1}{m_t m_b} \, (\bar{\xi}^{D}_{N,bs}\frac{V_{tb}}{V_{ts}^{*}}
+\bar{\xi}^{D}_{N,ss}) \, (\bar{\xi}^{U}_{N,tt}+\bar{\xi}^{U}_{N,tc}
\frac{V_{cb}}{V_{tb}}) G_{2}(y) \,\, ,
\label{CoeffH2}
\end{eqnarray}
where $x=m_t^2/m_W^2$ and $y=m_t^2/m_{H^{\pm}}^2$.
The functions $F_{1}(y)$, $F_{2}(y)$, $G_{1}(y)$ and $G_{2}(y)$ are given as
\begin{eqnarray}
F_{1}(y)&=& \frac{y(7-5y-8y^2)}{72 (y-1)^3}+\frac{y^2 (3y-2)}{12(y-1)^4}
\,
ln y \nonumber  \,\, , \\ 
F_{2}(y)&=& \frac{y(5y-3)}{12 (y-1)^2}+\frac{y(-3y+2)}{6(y-1)^3}\, ln y 
\nonumber  \,\, ,\\ 
G_{1}(y)&=& \frac{y(-y^2+5y+2)}{24 (y-1)^3}+\frac{-y^2} {4(y-1)^4} \, ln y
\nonumber  \,\, ,\\ 
G_{2}(y)&=& \frac{y(y-3)}{4 (y-1)^2}+\frac{y} {2(y-1)^3} \, ln y \,\, .
\label{F1G1}
\end{eqnarray}
In calculations we neglect the small contributions of the internal $u$ and $c$ quarks 
compared to one due to the internal $t$ quark.

For the initial values of the mentioned Wilson coefficients in the model III  
(eqs. (\ref{CoeffH}), (\ref{Coeffsm2}) and (\ref{CoeffH2})), we have 
\begin{eqnarray}
C_i^{(\prime )2HDM}(m_W)&=&C_i^{(\prime) SM}(m_W)+C_i^{(\prime) H}(m_W)\, .
\label{Coeff2HDM}
\end{eqnarray}
 Using these initial values, we can calculate the coefficients 
$C_{i}^{2HDM}$ and $C^{\prime 2HDM}_{i}$ at any lower 
scale with five quark effective theory where large logarithims can be 
summed using the renormalization group.
Since the strong interactions preserve chirality, the operators in eqs.
~(\ref{o7}, \ref{newop1}) can not mix with their chirality flipped 
counterparts eq. (\ref{newop2}) and the anomalous dimension matrices of two 
separate set of operators are the same and do not overlap.
With the above choosen initial values of Wilson coefficients,
their evaluations are similar to the SM case.

For completeness, note that, the operators $O_5$,$O_6$,$O_9$ and $O_{10}$ 
($O'_5$,$O'_6$, $O'_9$ and $O'_{10}$) give contributions to the 
matrix element of $b\rightarrow s\gamma$ and 
in the NDR scheme which we use here, the effective magnetic moment 
type Wilson coefficients are redefined as 
\begin{eqnarray}
C_{7}^{eff}(\mu)&=&C_{7}^{2HDM}(\mu)+ Q_d \, 
(C_{5}^{2HDM}(\mu) + N_c \, C_{6}^{2HDM}(\mu))\nonumber \, \, , \\
&+& Q_u\, (\frac{m_c}{m_b}\, C_{10}^{2HDM}(\mu) + N_c \, 
\frac{m_c}{m_b}\,C_{9}^{2HDM}(\mu))\nonumber \, \, , \\
C^{\prime eff}_7(\mu)&=& C^{\prime 2HDM}_7(\mu)+Q_{d}\, 
(C^{\prime 2HDM}_5(\mu) + N_c \, C^{\prime 2HDM}_6(\mu))\nonumber \\
&+& Q_u (\frac{m_c}{m_b}\, C_{10}^{\prime 2HDM}(\mu) + N_c \, 
\frac{m_c}{m_b}\,C_{9}^{\prime 2HDM}(\mu)) \, \, ,
\label{C7eff}
\end{eqnarray}
where $N_c$ is the color factor and $Q_u$ ($Q_d$) is the charge of 
up (down) quarks.
There is still another mixing in the operator set 
$O_{7},O_{8},O_{9},O_{10}$
($O'_{7},O'_{8},O'_{9},O'_{10}$) \cite{cho}  and we do not take into account 
since the initial values of the Wilson coefficients 
$C_{10}$ and $C^{\prime}_{10}$  are zero in our case.

The NLO corrected coefficients $C_{7}^{2HDM}(\mu)$ and
$C^{\prime 2HDM}_7(\mu)$  are given as
\begin{eqnarray}
C_{7}^{2HDM}(\mu)&=&C_{7}^{LO, 2HDM}(\mu) 
+\frac{\alpha_s (\mu)}{4\pi} C_7^{(1)\, 2HDM}(\mu) \nonumber \,\, , \\
C^{\prime 2HDM}_7(\mu)&=& C_{7}^{\prime LO, 2HDM}(\mu) + 
\frac{\alpha_s (\mu)}{4\pi} C_7^{\prime (1)\, 2HDM}(\mu)\,\, .
\label{renwils}
\end{eqnarray}
Here $\eta =\alpha_{s}(m_{W})/\alpha_{s}(\mu)$, $h_{i}$ and $a_{i}$ are 
the numbers which appear during the evaluation \cite{buras}.
The functions $C_{7}^{LO, 2HDM}(\mu)$ \cite{bmisiak} and 
$C_{7}^{\prime LO, 2HDM}(\mu$)are the leading order QCD corrected Wilson 
coefficients:
\begin{eqnarray} 
C_{7}^{\prime LO, 2HDM}(\mu)&=& \eta^{16/23} C^{\prime 2HDM}_7(m_{W})+
(8/3) (\eta^{14/23}-\eta^{16/23}) C^{\prime 2HDM}_8(m_{W}) \,\,
\label{LOwils}
\end{eqnarray}
and $C_7^{(1)\, 2HDM}(\mu)$ is the $\alpha_s$ correction to the leading
order result that its explicit form can be found in \cite{greub2,ciuchini2}.
$C_7^{\prime (1)\, 2HDM}(\mu)$ can be obtained  by  
replacing the Wilson coefficients in $C_7^{(1)\, 2HDM}(\mu)$ 
with their primed counterparts.
The NLO corrected coefficients $C_{5}^{2HDM}(\mu)$,
$C_{6}^{2HDM}(\mu)$ and  $C^{\prime 2HDM}_5(\mu)$, $C^{\prime 2HDM}_6(\mu)$ 
are numerically small at $m_b$ scale therefore we neglect them in our 
calculations.

Finally, the NLO QCD corrected $b\rightarrow s\gamma$ decay rate in model
III is obtained as 
\begin{eqnarray}
\Gamma (b\rightarrow s\gamma)=\frac{G_{F}^2 m_{b}^5}{32 \pi^4}
\alpha_{em} |V_{ts}^* V_{tb}|^2 (|C_{7}^{eff}(m_{b})|^2+
|C^{\prime eff}_7 (m_{b})|^2)\,\, , 
\label{decayrate}
\end{eqnarray} 
where $\alpha_{em}$ is the fine structure constant, and $m_b$ is b-quark mass.
$|C_{7}^{eff}(m_{b})|^2$ is given in \cite{greub2}
\begin{eqnarray}
|C_{7}^{eff}(m_{b})|^2&=&|D|^2+A+\frac{\delta_{\gamma}^{NP}}{m_b^2} 
|C_{7}^{LO, 2HDM}(m_b)|^2 \nonumber \\ &+& 
\frac{\delta_{c}^{NP}}{m_b^2} Re \{ \bigg (  C_{7}^{LO, 2HDM}(m_b) \bigg ) ^* 
\bigg ( C_{2}^{LO, 2HDM}(m_b)-\frac{1}{6} C_{1}^{LO, 2HDM}(m_b)\bigg )\} \,\, .
\label{C7tot}
\end{eqnarray}
The functions $D$ and A are \cite{greub2}
\begin{eqnarray}
D&=&C_{7}^{LO, 2HDM}(m_b)+\frac{\alpha_s}{4\pi} \bigg ( 
C_7^{(1)\, 2HDM}(\mu) + \sum _{i}^{8} C_{i}^{LO, 2HDM}(m_b) r_i 
-\frac{16}{3} C_{7}^{LO, 2HDM}(m_b) \bigg ) \nonumber \,\, , \\
A&=& \frac{\alpha_s(m_b)}{\pi} \sum_{i,j=1,i\leq j}^{8} 
Re\{ C_{i}^{LO, 2HDM}(m_b) \bigg ( C_{j}^{LO, 2HDM}(m_b) \bigg )^* 
f_{ij}\} \,\, .
\label{DA}
\end{eqnarray}
The explicite expressions for $f_{ij}$, $r_i$, $\delta^{NP}_{\gamma}$ and 
$\delta^{NP}_c$ can be found in \cite{greub2}. At this point we would like 
to note that the expressions for unprimed Wilson coefficients in our case 
can be obtained from the results in \cite{greub2} by the following 
replacements:
\begin{eqnarray}
|Y|^2 &\rightarrow& \frac{1}{m_{t}^2} \,
(\bar{\xi}^{U}_{N,tt}+\bar{\xi}^{U}_{N,tc}
\frac{V_{cs}^{*}}{V_{ts}^{*}}) \, (\bar{\xi}^{U}_{N,tt}+\bar{\xi}^{U}_{N,tc}
\frac{V_{cb}}{V_{tb}}) \nonumber \,\, , \\
XY &\rightarrow& \frac{1}{m_t m_b} \, (\bar{\xi}^{U}_{N,tt}+\bar{\xi}^{U}_{N,tc}
\frac{V_{cs}^{*}}{V_{ts}^{*}}) \, (\bar{\xi}^{D}_{N,bb}+\bar{\xi}^{D}_{N,sb}
\frac{V_{ts}}{V_{tb}})
\label{repl}
\end{eqnarray}

To obtain $|C_{7}^{\prime eff}(m_{b})|^2$, it is enough to use the primed 
Wilson coefficients at $m_W$ level \\ 
(eq. \ref{CoeffH2}) since the evaluation 
of $C_{7}^{\prime eff}(\mu)$ from $\mu=m_W$ to $\mu=m_b$ is the same as
that of $C_{7}^{eff}(\mu)$.

Note that, for model II (model I) $Y$ and $XY$ are 
\begin{eqnarray}
Y&=&1/tan\beta \,\, (1/tan\beta) \nonumber \,\,, \\
XY&=&1\,\, (-1/tan^2\beta) \,\, .
\end{eqnarray}
\section{Constraint analysis}
Now let us turn our attention to the constraint analysis.
Restrictions to the free parameters,
namely, the masses of the charged and neutral Higgs bosons
and the ratio of the v.e.v. of the two Higgs fields, 
denoted by $\tan\beta$ in the framework of model I and II, have been
predicted in series of works \cite{const}. 
Recently the constraint which connects masses of the charged Higgs
bosons, $m_{H^{\pm}}$ and $tan\beta$, is obtained by using the QCD corrected 
values in the LLO approximation and it is shown that the constraint region 
is sensitive to the renormalization scale, $\mu$ \cite{ahi}.  

Usually, the stronger restrictions to the new couplings are obtained from 
the analysis of the $\Delta F=2$ (here $F=K,B_d,D$) decays, the $\rho$ 
parameter and the $B\rightarrow X_s \gamma$ decay.  
In \cite{soni2}, all these processes have been analysed and two possible 
scenarios are obtained depending on the choice whether the constraint
from $R_b^{exp}$ is enforced or not.   
Although the new experimental results are near the SM result, 
$R_b^{SM} =0.2156$, the world average value for $R_b (=0.21656 \pm 0.00074)$ 
is still almost one standard deviations higher than the SM one.  
This brings the possibility that an enhancement to the SM value is still 
necessary to get the correct experimental one. Such an enhancement 
is reached under the conditions $\xi_{bb}^{D}\, >>1$ and 
$m_{H^\pm}\sim 400 \, GeV$ \cite{soni2}, where $\xi_{bb}^{D}$ is a model III parameter 
(see section 2) and $v$ is the vacuum expectation value of the Higgs field 
responsible for the generation of fermion masses. 

First, the constraints for the FC couplings from $\Delta F=2$ processes for 
the model III were investigated withouth QCD corrections, under the following
assumptions \cite{soni2}
\begin{eqnarray}
1.\,\,\,\, \lambda_{ij}&\sim& \lambda \nonumber \,\, ,\\
2.\,\,\,\, \lambda_{uj}&=&\lambda_{dj} <<  1\,,\,\, i,j=1,2,3 \nonumber\,\, ,
\end{eqnarray}
where $u(d)$ is up (down) quark and  $i,j$ are the generation numbers.

\hspace{4.75cm}3.\,\,\,\, case 2 and further assumption 
\begin{eqnarray}
\lambda_{bb}\,\,, \lambda_{sb} >> 1 \,\, and \,\, 
\lambda_{tt}\,\,, \lambda_{ct} << 1\,\, .
\label{lambda}
\end{eqnarray}
In the analysis, the ansatz 
\begin{eqnarray}
\xi_{ij}^{U D}=\lambda_{ij}\sqrt{\frac{m_i m_j}{v}}\,\, ,
\label{ansatz}
\end{eqnarray}
is used. Note that this ansatz coincides with the one proposed by 
Cheng and Sher.

Using the constraint coming from $R_{b}^{exp}$, the measurement 
$Br(B\rightarrow X_s \gamma )$, $\Delta F=2$ mixing and the result coming 
from the analysis of the $\rho$ parameter, the following restrictions are 
obtained \cite{soni2}:
\begin{eqnarray}
150 \, GeV \le m_{H^{\pm}} \le 200\, GeV \,\, , \nonumber \\
\lambda_{bb} >> 1 \,\, , \lambda_{tt} << 1 \,\, , \nonumber  \\
\lambda_{sb} >> 1  \,\, , \lambda_{ct} << 1 \,\, .
\label{lambda2}
\end{eqnarray}
Since the experimental results for $R_b^{exp}$ are still unclear,
we disregard the constraint coming from $R_b^{exp}$ and 
we analyse the restrictions for the couplings 
$\bar{\xi}^{U}_{N tt}$, $\bar{\xi}^{D}_{N bb}$ and $\bar{\xi}^{U}_{N tc}$ 
in the NLO aproximation, respecting the constraints due to the 
$\Delta F=2$ mixing, the $\rho$ parameter and using the measurement
by the CLEO \cite{cleo2} Collaboration: 
\begin{eqnarray}
Br (B\rightarrow X_{s}\gamma)&=&(3.15\pm 0.35\pm 0.32) \cdot 10^{-4} \,
\nonumber \,\, .
\label{branching}
\end{eqnarray}     
Here, we explain why we use only the CLEO data in our analysis
but not ALEPH one 
($Br (B\rightarrow X_{s}\gamma)=(3.11\pm 0.80\pm 0.72) \cdot 10^{-4} $). 
The ALEPH data has a larger error compared to CLEO 
data and it leads to a wide restriction region for $|C_7^{eff}|$, which 
includes the one coming from the CLEO data.
Therefore, the CLEO data allows us to get more stringest constraints on the 
model parameters. 

The idea in this calculation is to take $\bar{\xi}_{N tc} << 
\bar{\xi}^{U}_{N tt},  \bar{\xi}^{D}_{N bb}$ and 
$\bar{\xi}^{D}_{N ib} \sim 0\, , \bar{\xi}^{D}_{N ij}\sim 0$,
where the indices $i,j$ denote $d$ and $s$ quarks.
This choice permit us to neglect the neutral Higgs contributions because 
the Yukawa vertices are the combinations of $\bar{\xi}^{D}_{N ib}$ and 
$\bar{\xi}^{D}_{N ij}$.

To reduce the b-quark mass dependence let us consider the ratio 
\begin{eqnarray}
R&=&\frac{Br (B\rightarrow X_{s}\gamma)}
{Br (B\rightarrow X_{c} e \bar{\nu}_{e})}\nonumber \\
&=& \frac{|V_{ts}^{*} V_{tb}|^{2}}{|V_{cb}|^{2}}\frac{6 \alpha_{em}}
{\pi g(z) \kappa (z)} |C_{7}^{eff}|^{2} \, \, ,
\label{R}
\end{eqnarray}
where $g(z)$ is the phase space factor in semileptonic b-decay,
$\kappa (z)$ is the QCD correction to the semileptonic decay width
\cite{cabibbo},
\begin{eqnarray}
g(z)&=&1-8 z^{2}+8 z^{6}-z^{8}-24 z^{4} ln \,z \nonumber \,\, , \\
\kappa (z) &=&1-\frac{2 \alpha_s (m_b)}{3 \pi} \{
(\pi^2-\frac{31}{4})(1-z)+\frac{3}{2} \}- (0.25-0.18 (1-4 \frac{(1-z^2)^4}
{g(z)})\,\, , 
\label{gz}
\end{eqnarray}
and $z=m_{c}/m_{b}$.

Using the CLEO data and following the same procedure as given in \cite{ahi},
we reach the possible range for $|C_7^{eff}|$ as 
\begin{eqnarray}
0.257 \leq |C_{7}^{eff}| \leq 0.439 \, \, . 
\label{C7}
\end{eqnarray}

\begin{figure}[htb]
\vskip -0.15truein
\centering
\epsfxsize=5.0in
\leavevmode\epsffile{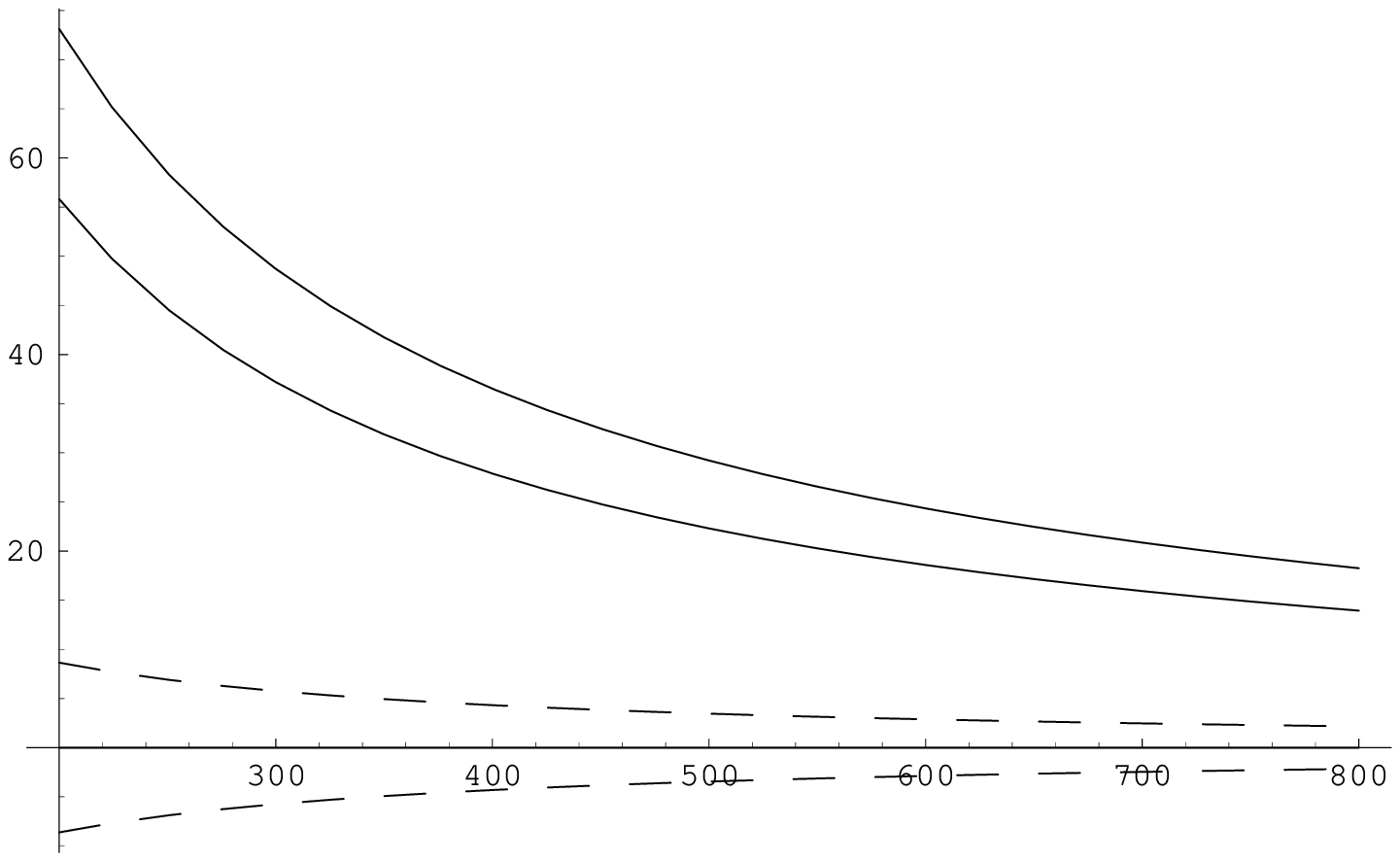}
\vskip -0.15truein
\vskip -12.4cm
\hspace{-12cm} $\bar{\xi}^{D}_{N bb}$  
\vskip 6cm
\hspace{11cm} $\bar{\xi}^{U}_{N tt}$  
\vskip 2cm 
\caption[]{$\bar{\xi}_{N tt}^{U}$ as a function of $\bar{\xi}_{N bb}^{D}$
for the fixed value of the charged Higgs boson mass $m_{H^{\pm}}=400\, GeV$ at 
$\mu=4.8 \, GeV$.  
Here the constraint region is lying in between solid (dashed) curves. The
solid (dashed) curves are the boundaries of the constraint region 
corresponding to $C_{7}^{eff} > 0$ \,\, ($C_{7}^{eff} < 0$) }
\label{tb45}
\end{figure}

In  fig.~(\ref{tb45}), we plot the parameter $\bar{\xi}_{N,tt}^{U}$ with 
respect to $\bar{\xi}_{N,bb}^{D}$ at $\mu=4.8 \,\, GeV$ and 
$m_{H^{\pm}}=400\, GeV$. We see, that there are two different restriction 
regions, where the upper one corresponds to the positive $C_{7}^{eff}$ value, 
however the lower one to the negative $C_{7}^{eff}$ value.
Increasing $\bar{\xi}_{N,bb}^{D}$ causes $|\bar{\xi}_{N,tt}^{U}|$ to 
decrease in both regions. With the given value of  
$\bar{\xi}_{N,bb}^{D} >> 1$, the condition 
$|r_{tb}|=|\frac{\bar{\xi}_{N,tt}^{U}}{\bar{\xi}_{N,bb}^D}| < 1$
is obtained. In the lower region it is possible that the ratio becomes
negative, i.e. $r_{tb}=\frac{\bar{\xi}_{N,tt}^{U}}{\bar{\xi}_{N,bb}^D} < 0$.
Further, increasing $m_{H^{\pm}}$ causes to increase $|r_{tb}|$ and the area
of the restriction region. 

\begin{figure}[htb]
\vskip -0.5truein
\centering
\epsfxsize=5.0in
\leavevmode\epsffile{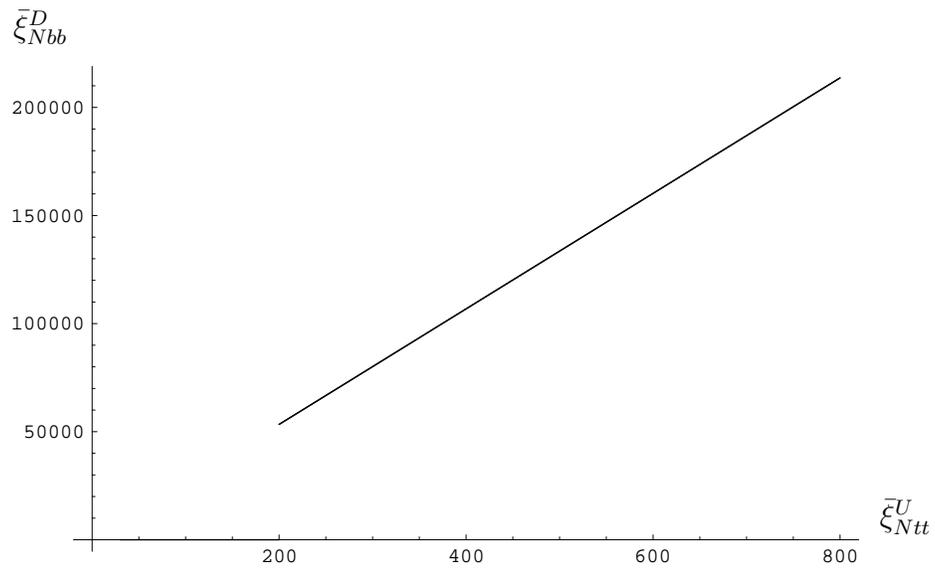}
\vskip -0.15truein
\vskip -12.4cm
\hspace{-12cm} $\bar{\xi}^{D}_{N bb}$  
\vskip 6cm
\hspace{11cm} $\bar{\xi}^{U}_{N tt}$  
\vskip 2cm 
\caption[]{Same as fig~1, but the third possible constraint region.}
\label{tb45b}
\end{figure}

Fig.~(\ref{tb45b}) is devoted the same dependence as in  fig.~(\ref{tb45}) 
and shows that the third region, which is almost a straight line, appears. 
In this region the ratio $r_{tb}>> 1$ and increases with increasing 
$m_{H^{\pm}}$ similar to the previous regions. 

\begin{figure}[htb]
\vskip -0.5truein
\centering
\epsfxsize=5.0in
\leavevmode\epsffile{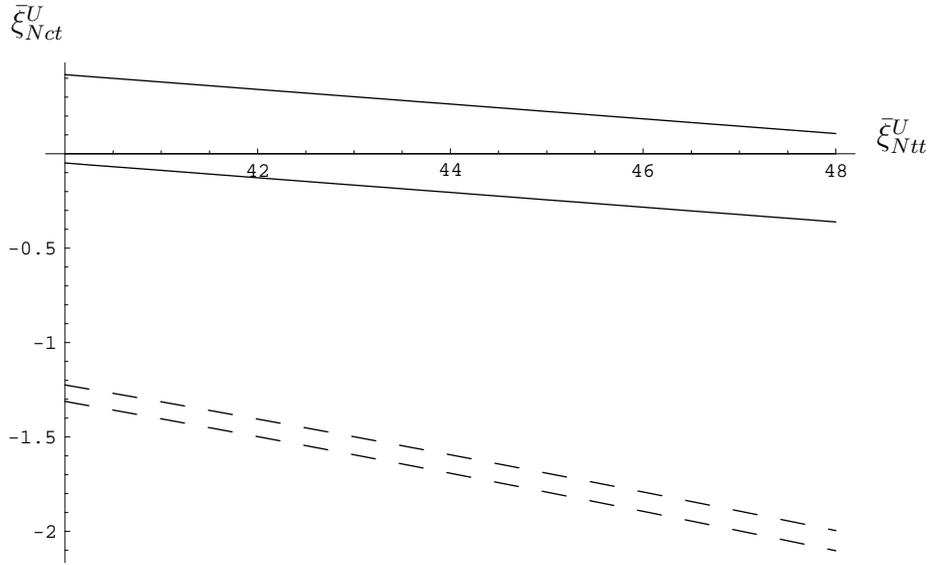}
\vskip -0.15truein
\vskip -12.4cm
\hspace{-12cm} $\bar{\xi}^{U}_{N ct}$  
\vskip 1cm
\hspace{11cm} $\bar{\xi}^{U}_{N tt}$  
\vskip 8cm 
\caption[]{$\bar{\xi}_{N tt}^{U}$ dependence of $\bar{\xi}_{N ct}^{U}$ 
for the fixed $\bar{\xi}_{N bb}^{U}=60 \,m_b$, at $\mu=4.8 \, GeV$
and $m_{H^{\pm}}=400\, GeV$.  
Here the constraint region is lying in between solid curves (dashed curves).
The solid (dashed) curves are the boundaries of the constraint region 
corresponding to $C_{7}^{eff} >0$\,\, ($C_{7}^{eff} <0$) }
\label{cttt564}
\end{figure}

Finally, we consider $\bar{\xi}_{N,tt}^{U}$ dependence of   
$\bar{\xi}_{N,tc}^{U}$, which is a neutral FC coupling.
In fig.~(\ref{cttt564}) we plot the $\bar{\xi}_{N,tt}^{U}$ dependence of 
$\bar{\xi}_{N,tc}^{U}$ for fixed $\bar{\xi}_{N,bb}^{D}=60 \, m_{b}$, at 
$\mu=4.8\, GeV$, and charged Higgs mass $m_{H^{\pm}}=400\, GeV$.
Here the selected region for $\bar{\xi}_{N,tt}^{U}$ is 
$40 \leq \bar{\xi}_{N,tt}^{U}\leq 48$. Increasing $\bar{\xi}_{N,tt}^{U}$ 
forces the ratio $r_{tc}=\frac{\bar{\xi}_{N,tc}^{U}}{\bar{\xi}_{N,tt}^{U}}$ 
to be negative. It is realized that the ratio $|r_{tc}|$ becomes smaller 
when $m_{H^{\pm}}$ is larger. 

Still there is a region in which $\bar{\xi}_{N,tc}^{U}$ is constrained 
for the possible large value of $\bar{\xi}_{N,tt}^{U}$, namely
$\bar{\xi}_{N,tt}^{U}=8.0 \, 10^4$ for $m_{H^{\pm}}=400\, GeV$: 
\begin{eqnarray}
-0.24 < \bar{\xi}_{N,tc}^{U} < 0.24 \nonumber \,\, , or \\
-3.26 < \bar{\xi}_{N,tc}^{U} < -3.19 \,\, , 
\label{ktc1}
\end{eqnarray}

In conclusion, we find  the constraints for the Yukawa couplings 
$\bar{\xi}_{N,tt}^{U}$, $\bar{\xi}_{N,bb}^{D}$ and $\bar{\xi}_{N,tc}^{U}$
using the CLEO measurement $Br(B\rightarrow X_{s}\gamma)$ and respecting the   
restrictions due to the $\Delta F=2$ mixing and the $\rho$ parameter
(see \cite{soni2} for details).
The constraints for the other parameters of the model III from the 
existing experimental results require more detailed new analysis.

\newpage

\end{document}